# Citation Pathways in the AI Era: Interpretive Knowledge Nodes, Citation Compression Layers, and the Measurement Boundary of Scholarly Impact

**Li Li[1],Yu Cao[2]**

[1]Hefei No.62 Middle School, Hefei 230031,Anhui, China
[2]Anhui NARI ZT Electric Co., Ltd., Hefei 230031, Anhui,China

**Abstract**

This article identifies "citation pathway" as a long-neglected analytical dimension in scientometrics. Traditional evaluation metrics focus on measuring citation counts while paying insufficient attention to the intermediate nodes through which knowledge flows from its original source to the citing author. Building on an analysis of the normative structure of current reference systems, this article introduces two new concepts: **Interpretive Knowledge Nodes (IKN)** — academic papers that provide structured reorganizations of classic works — and **Citation Compression Layers (CCL)** — the intermediate layers that emerge when such knowledge products acquire stable publication identities and enter formal citation networks at scale. The central proposition is that AI has not changed citation rules themselves but has transformed the cost structure of producing citable knowledge intermediaries. Under conditions of full compliance, the network position effects of citation pathways may become a salient variable affecting the validity of impact measurement. Through a thought experiment involving a hypothetical journal R and a parsimonious "Citation Gravity conceptual model," this article substantiates this proposition and discusses the measurement boundaries of scholarly impact indicators, as well as the institutional risk of incentive misalignment under extreme scenarios.

## 1 Introduction: A Neglected Dimension

Academic evaluation systems rely on a small set of core indicators to assess research quality, with citation analysis occupying a central position. Since Garfield introduced the impact factor[1], citation counts and their derivatives — the h-index, average citations per article, journal impact factors — have become de facto standards for academic hiring, promotion, and resource allocation. Criticisms of these indicators are well-rehearsed: the highly skewed distribution of citations renders means unrepresentative[2]; disciplinary differences in citation density undermine cross-field comparisons[3]; and citation manipulation — from excessive self-citation to "citation cartels" — continues to erode trust in these metrics[4].

Yet these criticisms share a common premise: they focus on the **compliance** of citing behavior or the **statistical properties** of metrics, without asking which intermediate nodes knowledge passes through on its journey from original source to citing author. In other words, the existing literature has addressed whether citations are legitimate, comparable, or manipulated, but has neglected a more fundamental dimension — the **citation pathway**.

This neglect is not accidental. Under historical conditions where knowledge mediation was expensive to produce, high-quality intermediary texts — review articles, textbooks, tutorial papers — were themselves scarce resources and could not systematically perturb the overall structure of citation networks. But as AI technology dramatically reduces the costs of knowledge reorganization and transcription, this premise is being undermined. This article thus poses the central question: **when the cost of producing knowledge intermediaries undergoes an AI-driven transformation, might fully compliant citation behavior generate unintended network-level effects?**

This article employs the methods of conceptual analysis and institutional thought experiment. Section 2 examines the normative structure of current reference systems and their observational blind spots. Section 3 introduces the two core concepts of IKN and CCL. Section 4 presents a thought experiment involving hypothetical journal R to demonstrate the mechanism of pathway effects. Section 5 constructs a Citation Gravity conceptual model to provide a formalized explanation. Section 6 introduces the concept of knowledge mediation elasticity and analyzes how AI has altered the cost-benefit equation. Section 7 discusses the boundaries of measurement validity and the institutional risks under extreme scenarios.

## 2 The Normative Structure of Reference Systems

To understand why citation pathways become a problem, one must first examine what current reference systems regulate — and what they leave unregulated.

### *2.1 The Explicit Constraint Layer*

The normative requirements imposed by academic publishing on references concentrate on three dimensions: **authenticity** (cited works must genuinely exist and be publicly available), **accessibility** (other researchers must be able to locate cited works through the bibliographic information provided), and **content relevance** (citations must bear substantive connections to the paper's argument). These requirements are codified in COPE's Core Practices[5], ICMJE's Recommendations[6], and the author guidelines of journals across disciplines, forming the "explicit constraint layer" of citation behavior — violations are treated as academic misconduct.

### *2.2 The Weak Constraint Layer*

However, on the question of "which level of source should be cited," the system lacks rigid provisions. The priority norm — attributing credit for a discovery to its earliest publisher — is a core principle of the sociology of science[7], but it functions as an academic cultural norm rather than an institutional rule. When a researcher needs to cite a classical concept, the system requires citing something authentic, accessible, and relevant, but does not mandate citing the original source rather than a clear review article. Academic writing guides typically "recommend" consulting original sources, but a recommendation is not a dimension of compliance review.

### *2.3 The Institutional Observation Blind Spot*

This means the effective unit of analysis for reference systems is the **individual paper**: reviewers and editors check whether a particular paper's citations are compliant, but no one checks whether the pathway structure of the entire citation network has undergone systematic migration. The system can detect a paper that fabricates references, but cannot detect ten thousand papers each independently and compliantly choosing the same intermediary text over the original source. This constitutes a structural observation blind spot — not because the system is poorly enforced, but because the monitoring scope of institutional design has never extended to network-level pathway changes.

## 3 Core Concepts: IKN and CCL

### *3.1 Interpretive Knowledge Nodes (IKN)*

This article defines an **Interpretive Knowledge Node (IKN)** as an academic paper that provides structured reorganization, semantic transcription, and format adaptation of a recognized classic work. The function of an IKN is to serve as an "interface layer" for knowledge dissemination: it re-presents a classic's core concepts, formulas, and argumentative structure in a clearer, more standardized form, thereby reducing the formatting burden and cognitive cost for subsequent researchers who wish to cite the work accurately.

At the level of institutional form, IKNs occupy exactly the same compliance position as existing legitimate intermediary texts — review articles, textbooks, and tutorial papers. All satisfy the four basic requirements for references: they genuinely exist, are publicly accessible, are relevant to the claims they support, and have been read by the citing author. A common objection is that review articles embody expert judgment and selection, whereas AI-assisted IKNs offer only mechanical reorganization, making the two fundamentally different. This article's response is that whether an IKN contains "judgment" is a question of **content quality**, not **institutional compliance**. The current reference system does not use "judgment" as a dimension of citation compliance review — indeed, reviews vary equally in quality, and a mediocre review is no different from an excellent IKN in terms of institutional compliance. If the system wishes to distinguish between "form-compliant but judgment-deficient" IKNs and expert-authored reviews, it would need to introduce a review dimension that currently does not exist — which is itself an illustration of the institutional observation blind spot identified in this article.

A key conceptual distinction is necessary here: an IKN is not a review article. The core function of a review article is the synthesis of a research field — its object is the body of literature within a domain, and its goal is to integrate research progress and map the distribution of viewpoints. The core function of an IKN, by contrast, is the transmission optimization of a **single** canonical knowledge object — its object is one specific classic work, and its goal is to reduce the cognitive and formatting cost for subsequent researchers who need to invoke that classic accurately. A review connects a research community; an IKN connects a classic node to application nodes. The two occupy the same institutional compliance position — both satisfy the four basic requirements — but differ structurally in cognitive function, operational object, and network role. This distinction is essential for understanding why IKNs produce a distinctive compression effect within citation networks.

### *3.2 Visibility and Scarcity of the Mediation Layer*

Not all knowledge mediation enters the citation network. Traditionally, classroom lectures, academic notes, and secondary interpretations in informal communication channels have played significant roles in knowledge circulation, but lacking DOIs or formal publication identities, they leave no trace in citation networks — they constitute an "invisible" mediation layer.

IKNs differ in one crucial respect: through formal publication, they acquire DOIs and stable academic identities, thereby becoming **visible nodes** in the citation network. This transition is significant because it transforms the knowledge mediation layer from "background infrastructure" of the citation network into a network element that can be measured and tracked. Historically, the academic community has defaulted to the assumption that high-quality intermediary texts are scarce — writing a competent review requires substantial expert time and domain judgment. This scarcity assumption is now being unsettled by new conditions of production.

### *3.3 Citation Compression Layer (CCL)*

When IKNs acquire stable publication identities and enter citation networks at scale, they form a new intermediate layer between original works (classics) and end citers (subsequent researchers). This article terms this phenomenon the **Citation Compression Layer (CCL)**. Its "compressive" nature manifests in two respects: cognitive compression — the IKN condenses a classic's complex argumentation into standardized formulations — and network compression — a substantial volume of citation relationships shifts from "researcher → classic" to a two-level structure of "researcher → IKN → classic," with the classic's direct connections in the network now mediated by the IKN. "Compression" here refers not to information loss alone, but to the transformation of high-dimensional knowledge structures into citation-efficient representations; the compressed IKN can still — and should — guide the reader back to the original work.

The core significance of the CCL lies not in whether it "blocks" direct citation of classics, but in that it transforms citation pathways from unobservable background conditions into **measurable network variables**. Once citation pathway becomes a variable, it enters the analytical purview of the evaluation system — even if the system has not yet recognized this variable's existence.

## 4 Hypothetical Journal R: A Thought Experiment

### *4.1 Logical Preliminaries*

Before introducing the thought experiment, a key logical structure must be clarified: the relationship between the independence of individual compliance and the holism of network migration. Consider the following extrapolation: ten thousand papers, each independently and fully compliantly, choose to cite an IKN rather than the original classic it reorganizes. Reviewers and editors examine each paper and find that every one satisfies the requirements of authenticity, accessibility, and content relevance — not a single violation. Yet after all ten thousand papers have made the same choice, a substantive shift in citation entry points has occurred at the systemic level: the classic's direct citation count stagnates or declines, while the IKN's rises continuously. The system can confirm compliance paper by paper but cannot aggregate ten thousand independent decisions into a structural network-level change for identification. This is the concrete manifestation of the institutional observation blind spot discussed in Section 2.

### *4.2 Construction of Hypothetical Journal R*

Now construct a concrete thought-experimental device: **hypothetical journal R** is an open-access journal whose articles primarily serve IKN-like intermediary functions — that is, structured reorganizations of highly-cited classics — standardized definitions of core concepts, formatted presentations of key formulas, and clear explications of argumentative structures (what it publishes); all articles satisfy the four reference system requirements of authentic existence, public accessibility, content relevance, and authorial reading, placing them in a state of full compliance (why it is legitimate); articles enter the citation network through DOIs and formal publication identities, becoming citable scholarly records (how they enter); each IKN provides equivalent citing functionality to the original classic but at a lower cognitive threshold and formatting adaptation cost, such that researchers, on cost-benefit grounds, preferentially cite the IKN over the original (why pathways shift).

### *4.3 Evolution and Outcome*

As journal R continues publication, two interrelated processes unfold simultaneously. At the micro level, increasing numbers of researchers find that citing an IKN is more efficient than consulting and citing the original — the IKN provides standardized formulations and bibliographic information that can be directly embedded into reference lists, without requiring researchers to extract and format this information themselves from the original. At the macro level, citation relationships redistribute within the citation network: substantial citation volume migrates from the original classic nodes to the IKN nodes.

R's impact factor rises accordingly — not because its papers possess cognitive originality, but because it occupies the intermediate position between classics and frontier research in the citation network. This is the direct manifestation of **betweenness centrality** at the level of evaluation metrics[8].

### *4.4 Methodological Note*

R is a purely analytical construct, not a description of any actual journal, and certainly not an advocacy of any publishing practice. The purpose of the thought experiment is not to promote but to test the logical boundary of an institutional norm: if an impact factor can change due to entirely legitimate position effects, then the scope within which impact factors can be unconditionally interpreted as indicators of scholarly contribution must be re-delimited.

## 5 The Citation Gravity Conceptual Model: A Mechanism for Position Effects

To give the above intuition greater precision, this section proposes a parsimonious formalized conceptual model.

### *5.1 The Implicit Assumption*

In current evaluation practice, citation count C is implicitly treated as a function of cognitive contribution Q: $C = f(Q)$. This is not a proposition anyone has explicitly asserted — no evaluation body has issued such a declaration — but it is the observable tacit logic of evaluative behavior: when a paper is highly cited, institutions and peers tend to interpret this as evidence that the paper possesses commensurate scholarly contribution, without asking what proportion of those citations stem from recognition of originality versus dissemination convenience.

### *5.2 The Revised Conceptual Model*

This article's revised conceptual model introduces a second variable — **Network Accessibility (A):**

$$\mathbf{C = f(Q, A)}$$

Here A describes, all else being equal, the probability that a paper will be accessed and cited by potential citers due to its network position and textual characteristics. Q and A are not orthogonal — a high-Q paper may also exhibit high A (through clear writing and broad dissemination) — but the two are conceptually separable.

*5.3 Decomposing A*

Network accessibility can be further decomposed into two sub-variables: **A = g(B, R)**.

**Betweenness Centrality (B)** derives from network science: in a citation network where nodes are papers and directed edges are citation relationships, a node's betweenness centrality measures the frequency with which it lies on shortest paths between other nodes[8]. An IKN situated between classics and large volumes of frontier research naturally possesses a high B value — not because it has "seized" a position in the network, but because the structure of the citation network endows it with a mediating function.

**Reading/Access Cost (R)** is a text-level variable encompassing cognitive threshold and formatting adaptation difficulty. An IKN that provides directly extractable standardized definitions and formatted reference entries has a substantially lower R value than an original classic that requires researchers to extract this information themselves.

*5.4 Explanatory Power and Testability*

Within this conceptual model, the behavioral logic of IKNs is clear: they hold Q roughly constant — they create no new knowledge — while increasing B by occupying advantageous network positions and reducing R by lowering textual barriers, thereby legitimately attaining higher C. The conceptual model decomposes "position effects" from a vague intuition into two independently measurable variables: B can be measured through network analysis modules of citation databases such as Web of Science or OpenAlex[9]; R can be assessed through readability metrics (e.g., Flesch Reading Ease) and text-mining approaches such as terminology coverage. Detailed operationalization is left to subsequent empirical research; the purpose here is solely to demonstrate the conceptual model's empirical testability.

## 6 AI, Knowledge Mediation Elasticity, and Cost Structure Transformation

*6.1 Knowledge Mediation Elasticity*

This article introduces the concept of **Knowledge Mediation Elasticity** to describe the responsiveness of the knowledge mediation layer's supply to demand. Before the large-scale deployment of AI technologies, knowledge mediation operated in a low-elasticity regime: writing a competent review or tutorial paper required substantial expert time with extremely high opportunity costs, so the supply of intermediary texts persistently lagged behind demand. Following the widespread adoption of AI-assisted writing tools, the cost of structurally reorganizing, semantically transcribing, and format-adapting classic works has dropped precipitously — work of equivalent quality no longer requires months of expert labor but can be completed at scale in a short time. Knowledge mediation elasticity has surged, and supply has shifted from a bottleneck to abundance.

*6.2 Core Mechanism: Rules Unchanged, Costs Transformed*

The central proposition of this article can now be stated with precision: AI has not modified any rule of the reference system. Authenticity, accessibility, content relevance, authorial reading — the four requirements remain as they were. What AI has changed is not the compliance standard but the **cost** of producing compliant knowledge intermediaries. Before AI intervention, the production cost of an IKN exceeded its potential citation returns (citing an IKN alone could hardly generate high impact), so there was no economic incentive for mass IKN production. After AI reduced production costs, this cost-benefit equation underwent a directional shift: IKNs can be produced at low cost, high quality, and large scale; scaled IKN production aggregates citation flows into journal-level impact; journal impact in turn incentivizes further IKN production — a positive feedback loop is thereby established.

*6.3 "AI Does Not Create New Rules; It Dramatically Increases the Supply of Actors That Can Operate Within Existing Ones"*

The institutional space has long existed — review articles and textbooks have always been legitimate citation targets, and the formation of a CCL has always been logically possible. But in the past, there were not enough "vehicles" passing through this institutional space: the supply of high-quality intermediary texts was insufficient to form an observable structural intermediate layer in citation networks. AI's role is not to create new institutional pathways but to mass-produce vehicles capable of traversing existing ones. The system was designed without considering the parameter value "mediation layer production cost approaching zero," because this lay beyond the technological imagination at the time of design. The system now faces a scenario it was never designed to handle.

## 7 Discussion and Conclusion

*7.1 Distinguishing Dissemination Value from Originality Value*

A necessary conceptual clarification is in order. Interpretive Knowledge Nodes possess genuine scholarly dissemination value: they lower cognitive thresholds, facilitate cross-disciplinary understanding, and enhance research efficiency. Such contributions deserve recognition within the academic ecosystem, and this article's argument is in no way opposed to knowledge intermediaries receiving citations.

However, dissemination value and originality value are distinct evaluative dimensions. An IKN receiving 1,000 citations for its dissemination value is entirely legitimate at the level of scholarly communication; but interpreting those 1,000 citations directly as evidence of "commensurate original contribution" constitutes a category confusion. Current citation indicators cannot institutionally distinguish between these two types of value — not through any failure of indicator design, but because under historical conditions of high mediation production costs, dissemination value and originality value were reasonably correlated, and conflating the two dimensions imposed limited costs on evaluative accuracy. This article's concern is whether, as AI reduces mediation production costs, the correlation between these two dimensions will decline significantly, such that the long-standing institutional conflation generates substantially greater measurement bias.

*7.2 Extreme Scenarios of Pathway Effects*

If citation pathway effects are significantly amplified by the surge in knowledge mediation supply elasticity, the academic evaluation system may face an institutional incentive shift. Under an extreme scenario, the evaluative signal may shift from "who made the original contribution" to "who occupies the optimal network position." Researchers will be incentivized not only to produce original knowledge but also to produce or facilitate the publication of Interpretive Knowledge Nodes in order to capture betweenness centrality within citation networks — and the latter may offer a superior input-output ratio, because reducing the access cost of existing knowledge is more predictable and more scalable than creating genuinely new knowledge.

This article does not predict that this scenario will necessarily materialize. It merely points out that the current evaluation system was designed without treating citation pathways as a variable, and therefore lacks mechanisms to prevent such an incentive shift. Under historical conditions of low knowledge mediation elasticity, this missing mechanism did not affect the system's overall incentive direction. Under AI conditions, whether this missing mechanism can still preserve the system's incentive validity is an empirical question warranting sustained scrutiny.

### *7.3 Theoretical Contributions*

This article's theoretical contributions operate at three levels. First, it identifies "citation pathway" as a long-neglected dimension within the analytical framework of scientometrics, extending scholarly critique from the compliance of citing behavior to the pathway structure of citation networks. Second, it proposes the conceptual tools of IKN, CCL, and knowledge mediation elasticity, providing a terminological foundation for describing and analyzing citation pathway effects. Third, it constructs a Citation Gravity conceptual model ($C = f(Q, A)$; $A = g(B, R)$) that decomposes "network position effects" from a vague intuition into variables that can be independently defined and measured, offering an actionable conceptual architecture for subsequent empirical research.

### *7.4 Limitations and Outlook*

This article is a work of conceptual analysis and institutional thought experiment. Its arguments rest on logical reconstruction of institutional norms and extrapolation from technological trends, rather than on quantitative analysis of actual citation data. Its core contribution is therefore the conceptual identification and mechanism characterization of a potential problem, not empirical proof that the problem has already materialized. Future empirical research should advance along at least two fronts: first, measuring actual changes in knowledge mediation elasticity within specific disciplines; second, testing the magnitude of pathway effects in real citation networks and their actual degree of perturbation on indicators such as impact factors. Only by joining the conceptual framework with empirical data can we assess the extent to which the logical possibilities identified in this article have already become substantive institutional challenges.